\documentclass[12pt]{iopart}
\usepackage{iopams}  
\input epsf
\expandafter\let\csname equation*\endcsname\relax
  \expandafter\let\csname endequation*\endcsname\relax
  \usepackage[fleqn]{amsmath}
\usepackage{amsfonts}
\usepackage{amssymb}
\usepackage{amsopn}
\usepackage{epsfig}
\usepackage{graphics,psfrag,rotating}
\usepackage{graphicx}
\usepackage{dcolumn}
\usepackage{bm}
\usepackage{epstopdf}
\usepackage{color}
\usepackage[usenames,dvipsnames,svgnames]{xcolor}
\usepackage[colorlinks=true,
            linkcolor=red,
            urlcolor=gray,
            citecolor=blue]{hyperref}
  \usepackage{hyperref}

\def \D {\tilde{\nabla}}

\def \ep {\varepsilon}
\def\rd {\displaystyle{\cdot}}

\def\div {\mbox{div}\,}
\def\md{\mathcal{D}}
\def\mz{\mathcal{Z}}

\def \ts {\textstyle}
\def\om{\omega}

\def\3nab{\tilde{\nabla}}

\def\la {\langle}
\def\ra {\rangle}

\def\div{\mbox{div}}
\def\tl{\tilde}
\def\hsp5{\hspace{5mm}}
\newcommand{\sfrac}[2]{{\textstyle{#1\over#2}}}
\def\case#1/#2{\textstyle\frac{#1}{#2}}
\def\rd {\displaystyle{\cdot}}
\def\ts {\textstyle}
\def\ber {\begin{eqnarray}}
\def\eer {\end{eqnarray}}
\def\bea {\begin{eqnarray}}
\def\eea {\end{eqnarray}}

\def\ts {\textstyle}
\def\bc {\begin{center}}
\def\ec {\end{center}}
\def\case#1/#2{\frac{#1}{#2}}

\newcommand{\bw}{\begin{widetext}}
\newcommand{\ew}{\end{widetext}}
\newcommand{\nn}{\nonumber\\}

\newcommand{\be}{\begin{equation}}
\newcommand{\bse}{\begin{subequation}}
\newcommand{\ese}{\end{subequation}}
\newcommand{\ee}{\end{equation}}
\newcommand{\eei}{\end{eqnarray}\indent\indent}
\newcommand{\ba}{\begin{array}}
\newcommand{\ea}{\end{array}}
\newcommand{\bal}{\begin{eqnarray}}
\newcommand{\eal}{\end{eqnarray}}

\newcommand{\hs}{\,-\,}

\newcommand{\car}{{\cal R}}

\def\case#1/#2{\textstyle\frac{#1}{#2} }
\newcommand{\nb}{\nabla}

\newcommand{\bver}{\begin{verbatim}}
\newcommand{\ever}{\end{verbatim}}

\begin{document}
\title{Anti-Newtonian cosmologies in $f(R)$ gravity}
\author{Amare Abebe \footnote{
amare.abbebe@gmail.com}}
\address{Department of Physics, North-West University,  Mafikeng 2735, South Africa.}

\begin{abstract}

 In this paper, we investigate a class of  perfect-fluid  ``anti-Newtonian'' cosmological models  in the context of $f(R)$ gravity. In particular, we study the integrability conditions of such gravity models  using covariant consistency analysis formalisms. We show that, unlike the results in General Relativity, anti-Newtonian cosmologies are not silent models and that they can exist subject to the solution of an integrability condition equation we derive. We also present the set of evolution equations governing the linear perturbations of matter, expansion and Ricci scalar for this class of models. \end{abstract}
  
 \maketitle


\section{Introduction}
Irrotational dust spacetimes in General Relativity (GR) have been studied as potential models for the description of gravitational collapse \cite{bert94, maartens98, wyll06} and late-time cosmic structure \cite{maartens94,ellis85id, maartens98}. These are models of spacetime characterised by vanishing isotropic pressure $p$ and vorticity $\omega_a$ but a positive-definite energy density $\mu\;,$ as well as a generally non-vanishing shear (an  exception being FLRW models) and non-vanishing locally free gravitational field.  The locally free gravitational field in such spacetimes is covariantly described by the gravito-electric (GE) and gravito-magnetic (GM) tensors, $E_{ab}$ and $H_{ab}\;,$ respectively \cite{maartens98}. Whereas $E_{ab}$ corresponds to the tidal tensor in the classical Newtonian gravitational theory, $H_{ab}$ is responsible for gravitational radiation in relativistic theories. $H_{ab}$, therefore, does not have a Newtonian analogue  and is taken to vanish identically for the limiting case. Such limiting cases of irrotational dust universes with vanishing GM part of the Weyl tensor (hence purely gravito-electric)  have come to be known as {\it Newtonian-like} universes. On the other hand, it is at least logically possible to consider those irrotational dust universes with purely gravito-magnetic Weyl tensor, i.e., vanishing EM field. These classes of models have been referred to as {\it anti-Newtonian} universes \cite{maartens98}.

Detailed consistency analyses of the propagation and constraint equations (arising from the Ricci identity for the fluid $4$-velocity and the Bianchi equations) for these models have been made over the years. In particular, the inconsistencies and resulting integrability conditions arising from  imposing external restrictions such as turning off the shear and/or  GM  (GE) tensors have been pointed out in \cite{maartens97, elst97, maartens98, elst98, wyll06}.

Maartens {\it et al} \cite{maartens98} showed that anti-Newtonian universes  (in GR) suffer from severe integrability conditions and conjectured that there are no anti-Newtonian spacetimes that are linearized perturbations of Friedman-Lema\^{i}tre-Robertson-Walker (FLRW) universes. This conjecture was later proved by Wylleman \cite{wyll06}.
In the present work, we  show that this is not necessarily true in fourth-order gravitational theories where the anisotropic pressure and heat-flux terms are generally not vanishing. 

Fourth-order theories of gravity have attracted huge attention recently, mainly because of their potential candidacy for dark energy-inspired cosmic acceleration and early universe inflation. These are a class of higher-order gravity models that attempt to address the shortcomings of GR in the infrared (IR) and ultraviolet (UV) ranges, i.e., very low and very high energy scales \cite{capozziello11extended, modesto12, biswas12,  clifton12}. They are generally obtained by including higher order curvature invariants in the
Einstein-Hilbert action, or by making the action non-linear in the Ricci curvature $R$ and$/$or contain terms involving combinations of derivatives of $R$, in which case the models are known as {\it  $f(R)$ theories of gravity}.

First proposed by Buchdal \cite{buchdal70}, $f(R)$ theories gained more popularity after further developments by Starobinsky \cite{staro80}  and later  following the realization of the  discrepancy between theory and observation \cite{carroll04,faraoni08, sotiriou07,biswas12, nojiri2011unified, nojiri06, nojiri07, sriva08, capoz06, de2010f, magnano87}. Since  $f(R)$ gravity theories  do generally have  enough freedom to produce any kind of cosmic background evolution history or  to explain some aspect of cosmological phenomena or another by a suitable choice of the defining action functional, a consistency analysis of the field equations describing those phenomena seems in order. This paper is an attempt in that direction: we not only aim at showing what kind of generalised consistency relations one can make by putting on purely gravito-magnetic restrictions to the Weyl tensor in $f(R)$ gravity theories, but also initiate discussions on the kind of cosmological analysis we can make (such as gravitational collapse and late time cosmological structure formation) based on viable anti-Newtonian solutions.

The paper is organized as follows:  in Sec.\ref{covsec} we give a summary of the covariant description and the necessary covariant equations. In Sec.\ref{linsec} we present the general linearized field equations and specialize to their anti-Newtonian limiting cases. We study the integrability conditions required to make the anti-Newtonian equations consistent in Sec.\ref{intsec}
 and show that there are anti-Newtonian solutions in $f(R)$ gravity whose linearized covariant equations are consistent.  Sec.\ref{pertsec} deals with the formulation of the  covariant density, expansion and curvature  perturbations.
Finally in Sec.\ref{concsec} we discuss the results and give an outline of future work.

Natural units ($\hbar=c=k_{B}=8\pi G=1$)
will be used throughout this paper, and Latin indices run from 0 to 3.
The symbols $\nabla$, $\D$ and the overdot $^{.}$ represent the usual covariant derivative, the spatial covariant derivative, and differentiation with respect to cosmic time.  We use the
$(-,+,+,+)$ signature and the Riemann tensor is defined by
\begin{eqnarray}
R^{a}_{bcd}=\Gamma^a_{bd,c}-\Gamma^a_{bc,d}+ \Gamma^e_{bd}\Gamma^a_{ce}-
\Gamma^f_{bc}\Gamma^a_{df}\;,
\end{eqnarray}
where the $\Gamma^a_{bd}$ are the Christoffel symbols (i.e., symmetric in
the lower indices), defined by
\begin{equation}
\Gamma^a_{bd}=\frac{1}{2}g^{ae}
\left(g_{be,d}+g_{ed,b}-g_{bd,e}\right)\;.
\end{equation}
The Ricci tensor is obtained by contracting the {\em first} and the
{\em third} indices of the Riemann tensor:
\begin{equation}\label{Ricci}
R_{ab}=g^{cd}R_{cadb}\;.
\end{equation}
Unless otherwise stated, primes $^{'}$ etc  are shorthands for derivatives with respect to the Ricci scalar
\be
R=R^{a}{}_{a}\;
\ee
and $f$ is used as a shorthand for $f(R)$.
\section{Covariant Description}\label{covsec}

The action of a generalized fourth-order gravity is given by 
\be
{\cal A}= \sfrac12 \int d^4x\sqrt{-g}\left[f(R)+2{\cal L}_m\right]\;,
\label{action}
\ee
where ${\cal L}_m$ represents the matter contribution to the Lagrangian,
and the equations
 \be
 G_{ab}=\tl T^{m}_{ab}+T^{R}_{ab}\equiv T_{ab}\;,
 \ee
 generalize Einstein's field equations, where 
 \be\label{emt}
\tl T^{m}_{ab}=\frac{T^{m}_{ab}}{f'}\;,~~~~
T^{R}_{ab}=\frac{1}{f'}\left[\sfrac{1}{2}(f-Rf')g_{ab}+\nb_{b}\nb_{a}f'-g_{ab}\nb_{c}
\nb^{c}f' \right]\;; f'\equiv df/dR\,.
\ee
The matter energy-momentum tensor is given by 
\be T^{m}_{ab} = \mu_{m}u_{a}u_{b} + p_{m}h_{ab}+ q^{m}_{a}u_{b}+ q^{m}_{b}u_{a}+\pi^{m}_{ab}\;,\ee
where  $\mu_{m}$, $p_{m}$, $q^{m}_{a}$ and $\pi^m_{ab}$ denote \footnote{Throughout this paper, we will use the $m$ and $R$ subscripts and superscripts  interchangeably only for convenience, and hence should not be confused with raising and lowering of indices.} the standard matter energy density, 
pressure, heat flux and anisotropic pressure respectively.  Here  $u^a$  is the $4$\hs velocity of fundamental observers:
\be u^{a} = \frac{dx^{a}}{dt}\;,\ee
whereas  \be h_{ab}=g_{ab}+u_au_b\ee
is the projection tensor into the tangent 3-spaces orthogonal to $u^a$.

The {\it total} thermodynamics of the composite matter-curvature fluid is then defined by
\be\label{totaltherm}
\mu\equiv\frac{\mu_{m}}{f'}+\mu_{R}\;,~~~\;p\equiv\frac{p_{m}}{f'}+p_{R}\;,~~~
q_{a}\equiv \frac{q^{m}_{a}}{f'}+q^{R}_{a}\;,~~~\;\pi_{ab}\equiv\frac{\pi^{m}_{ab}}{f'}+\pi^{R}_{ab}\;,
\ee
where $\mu^{R}$, etc. are thermodynamical quantities of the {\it curvature fluid}, the energy-momentum tensor of which is $T^{R}_{ab}$ as defined in Eqn \eref{emt}.

In the standard  $1+3$-covariant approach, two derivatives are defined: the 4-velocity vector $ u^{a} $ is used to define the 
\textit{covariant time derivative} (denoted by a dot) for any tensor 
${S}^{a..b}_{c..d} $ along an observer's worldlines:
\be
\dot{S}^{a..b}_{c..d}{} = u^{e} \nb_{e} {S}^{a..b}_{c..d}~,
\ee
and the tensor $ h_{ab} $ is used to define the fully orthogonally 
\textit{projected covariant derivative} $\tl\nb$ for any tensor ${S}^{a..b}_{c..d} $:
\be
\tl\nb_{e}S^{a..b}_{c..d}{} = h^a_f h^p_c...h^b_g h^q_d 
h^r_e \nb_{r} {S}^{f..g}_{p..q}\;,
\ee
with total projection on all the free indices. 
Angled brackets denote orthogonal projections of vectors and the
orthogonally \textit{projected symmetric trace-free} PSTF part of tensors is defined as
\be
V^{\langle a \rangle} = h^{a}_{b}V^{b}~, ~ S^{\langle ab \rangle} = \left[ h^{(a}_c {} h^{b)}_d 
- \sfrac{1}{3} h^{ab}h_{cd}\right] S^{cd}\;.
\label{PSTF}
\ee
The volume element for the 3-restspaces orthogonal to $u^a$ is defined by \cite{Ellis98, Abebe2011, betschart}:
\be
\ep_{abc}=u^{d}\eta_{dabc}=-\sqrt{|g|}\delta^0_{\left[ a \right. }\delta^1_b\delta^2_c\delta^3_{\left. d \right] }u^d\Rightarrow \ep_{abc}=\ep_{[abc]},~\ep_{abc}u^{c}=0,
\ee
where $\eta_{abcd}$ is the 4-dimensional volume element such that 
\be
\eta_{abcd}=\eta_{[abcd]}=2\ep_{ab[c}u_{d]}-2u_{[a}\ep_{b]cd}.
\ee
In particular, $\eta_{0123}=\sqrt{|det ~g_{ab}|}$.\\
$\ep_{abc}$ satisfies the following identities:
\begin{align}
&\ep^{abc}\ep_{def}=3!h^{[a}{}_{d}h^{b}{}_{e}h^{c]}{}_{f}\;,\\
&\ep^{abc}\ep_{cef}=2!h^{[a}{}_{e}h^{b]}{}_{f}\;,\\
&\ep^{abc}\ep_{bcf}=2!h^{a}{}_{f}\;,\\
&\ep^{abc}\ep_{abc}=3!\;.
\end{align}
The covariant spatial divergence and curl of vectors and tensors  are defined as \cite{maartens97,maart97, maartens98}
\ber
&& \div V=\D^aV_a\,,~~~~~~(\div S)_a=\D^bS_{ab}\,, \\
&& curl V_a=\ep_{abc}\D^bV^c\,,~~ curl S_{ab}=\ep_{cd(a}\D^cS_{b)}{}^d \,.
\eer

The covariant derivative of the timelike vector $u^a$ is decomposed into its
irreducible parts as
\be
\nb_au_b=-A_au_b+\sfrac13h_{ab}\Theta+\sigma_{ab}+\ep_{a b c}\omega^c,
\ee
where $A_a=\dot{u}_a$ is the acceleration, $\Theta=\tl\nb_au^a$ is the expansion, 
$\sigma_{ab}=\tl\nb_{\langle a}u_{b \rangle}$ is the shear tensor and $\omega^{a}=\ep^{a b c}\tl\nb_bu_c$ 
is the vorticity vector. 
The trace-free part of the Riemann tensor defines the {\it Weyl  conformal curvature tensor} $C_{abcd}$   \cite{Ellis98, betschart}
\be
C^{ab}{}_{cd}=R^{ab}{}_{cd}-2g^{[a}{}_{[c}R^{b]}{}_{d]}+\frac{R}{3}g^{[a}{}_{[c}g^{b]}{}_{d]}.
\ee
Because this tensor is trace-free,
\be
C^{a}{}_{bad}=0
\ee
and can be split into its ``electric'' and ``magnetic'' parts, $E_{ab}$
and $H_{ab}$ respectively given by
\be
E_{ab}\equiv C_{agbh}u^{g}u^{h},~~~~~~~H_{ab}=\sfrac{1}{2}\eta_{ae}{}^{gh}C_{ghbd}u^{e}u^{d}.
\ee
These tensors are each symmetric and trace-free in the local rest frame  of $u^{a}$:
\begin{align}
&E_{ab}=E_{(ab)},~~~~ H_{ab}=H_{(ab)},\nn
&E^{a}{}_{a}=0,~~~~~~~~H^{a}{}_{a}=0,\nn
&E_{ab}u^{b}=0,~~~~~~H_{ab}u^{b}=0.
\end{align}
Using these tensors the Weyl tensor can be rewritten as
\be
C_{abcd}=(\eta_{abpq}\eta_{cdrs}+g_{abpq}g_{cdrs})u^{p}u^{r}E^{qs}+(\eta_{abpq}g_{cdrs}+g_{abpq}\eta_{cdrs})u^{p}u^{r}H^{qs},
\ee
where 
\be
g_{abcd}=g_{ac}g_{bd}-g_{ad}g_{bc}.
\ee
$E_{ab}$ and $H_{ab}$  represent the free gravitational field, enabling gravitational action at a distance (tidal forces and gravitational waves), and influence the motion of matter and radiation through the geodesic deviation for timelike and null vectors respectively \cite{Ellis98}. As mentioned earlier, the magnetic part does not have a Newtonian analogue but the electric part  does, and can be given by
\be
E_{\alpha\beta}=\psi_{,\alpha\beta}-\sfrac{1}{3}h_{\alpha\beta}\psi^{\delta}{}_{,\delta}\;,
\ee
where $\psi$ represents  the Newtonian gravitational potential \cite{DPhD}.

It can be shown that for any scalar $\phi$ \cite{Abebe2011}
\ber
[\tl\nb_a\tl\nb_b-\tl\nb_b\tl\nb_a]\phi&=&2\ep_{a b c}\omega^c\dot \phi \;, \nonumber\\ 
\ep^{a b c}\tl\nb_b\tl\nb_c \phi&=&2\omega^a \dot \phi\;;
\label{C1}
\eer
and to linear order, we have 
\ber\label{C2}
&&[\tl\nb^a\tl\nb_b\tl\nb_a-\tl\nb_b\tl\nb^2]\phi=\sfrac{1}{3}\tl{R}\tl\nb_{b}\phi\;,\\
&&[\tl\nb^2\tl\nb_b-\tl\nb_b\tl\nb^2]\phi=\sfrac{1}{3}\tl{R}\tl\nb_{b}\phi+2\ep_{dbc}\tl\nb^d(\omega^c\dot \phi)\;,
\label{C3}
\eer
where $\tl{R}=\frac{6K}{a^{2}}=2\left(\mu-\frac13\Theta^2\right)$ is the 3-curvature scalar, $K=-1, 0$ or $1$ and $a=a(t)$ is the cosmological scale factor.
Also for any first order 3-vector $V^a=V^{\langle a \rangle}$, we have
\ber
[\tl\nb^a\tl\nb_b-\tl\nb_b\tl\nb^a]V_a&=&\sfrac13\tl{R}h^a_{\left[a\right. }
V_{\left. b \right]}\;,
\label{C4}
\eer
\ber
h^{a}_{c}h^{d}_{b}(\tl\nb_dV^c)\dot{}=\tl\nb_b\dot{V^{\langle a \rangle}}-\sfrac{1}{3}\Theta \tl\nb_bV^a,
\label{C6}
\eer
\ber
h^{a}_{c}(\tl\nb^2V^c)\dot{}=\tl\nb_b(\tl\nb^{\langle b}V^{a \rangle})\dot{}-\sfrac{1}{3}\Theta \tl\nb^2V^a.
\label{C5}
\eer

\section{Linearized Field Equations }\label{linsec}
 According to  the Stewart-Walker 
lemma \cite{SW74}, those cosmological quantities that vanish in the background spacetime are considered to be first order and  gauge-invariant in the covariant description.
In $f(R)$ gravity, the linearized thermodynamic quantities for the {\it curvature fluid}  are given by
\ber
&&\label{mur}\mu_{R}=\frac{1}{f'}\left[\sfrac{1}{2}(Rf'-f)-\Theta f'' \dot{R}+ f''\tilde{\nabla}^{2}R \right]\;,\\
&&\label{pr}p_{R}=\frac{1}{f'}\left[\sfrac{1}{2}(f-Rf')+f''\ddot{R}+f'''\dot{R}^{2}+\sfrac{2}{3}\left( \Theta f''\dot{R}-f''\tilde{\nabla}^{2}R \right) \right]\;,\\
&&\label{qar}q^{R}_{a}=-\frac{1}{f'}\left[f'''\dot{R}\tilde{\nabla}_{a}R +f''\tilde{\nabla}_{a}\dot{R}-\sfrac{1}{3}f''\Theta \tilde{\nabla}_{a}R \right]\;,\\
&&\label{pir} \pi^{R}_{ab}=\frac{f''}{f'}\left[\tilde{\nabla}_{\langle a}\tilde{\nabla}_{b\rangle}R-\sigma_{ab}\dot{R}\right]\;.
\eer
It is worth mentioning here that these quantities automatically vanish in GR, {\it i.e,} when  $f(R)=R$.
 
By covariantly $1+3$-splitting the Bianchi identities
\be\label{bi}
\nb_{[a}R_{bc]d}{}^{e}=0\;,
\ee
 and the Ricci identity
\be\label{ricci}
(\nb_{a}\nb_{b}-\nb_{b}\nb_{a})u_{c}=R_{abc}{}^{d}u_{d}\;
\ee
 for the total fluid 4-velocity $u^{a}$, we obtain the following propagation and constraint equations \cite{maartens98, carloni08, carlonireview}:
\ber
&&\label{mue}\dot{\mu}_{m}=-(\mu_{m}+p_{m})\Theta-\tl\nb^{a}q^{m}_{a}\;,\\
&&\dot{\mu}_{R}=-(\mu_{R}+p_{R})\Theta+\frac{\mu_{m}f''}{f'^{2}}\dot{R}-\D^{a}q^{R}_{a}\;,\\
&&\label{ray}\dot{\Theta}=-\sfrac13 \Theta^2-\sfrac12(\mu+3p)+\tl\nb_aA^a\;,\\
&&\label{qe}\dot{q}^{m}_{a}=-\sfrac{4}{3}\Theta q^{m}_{a}-\mu_{m}A_{a}\;,\\
&&\label{qer}\dot{q}^{R}_{a}=-\sfrac{4}{3}\Theta q^{R}_{a}+\frac{\mu_{m}f''}{f'^{2}}\D_{a}R-\D_{a}p_{R}-\D^{b}\pi^{R}_{ab}\;,\\
&&\label{propom}\dot{\omega}_{a}=-\sfrac23\Theta\omega_{a}-\sfrac{1}{2}\ep_{abc}\tl\nb^{b}A^{c}\;,\\
&&\label{sig}\dot{\sigma}_{ab}=-\sfrac{2}{3}\Theta\sigma_{ab}-E_{ab}+\sfrac{1}{2}\pi_{ab}+\tl\nb_{\la a}A_{b\ra}\;,\\
&&\label{gep}\dot{E}_{ab}+\sfrac{1}{2}\dot{\pi}_{ab}=\ep_{cd\langle a}\tl\nb^{c}H_{b\rangle }^{d}-\Theta E_{ab}-\sfrac{1}{2}\left(\mu+p\right)\sigma_{ab}
-\sfrac{1}{2}\tl\nb_{\langle a}q_{b\rangle}-\sfrac{1}{6}\Theta\pi_{ab}\;,\\
&&\label{gmp}\dot{H}_{ab} =-\Theta H_{ab}-\ep_{cd\langle a}\tl\nb^{c}E_{b\rangle }^{d}+
\sfrac{1}{2}\ep_{cd\langle a}\tl\nb^{c}\pi^{~d}_{b\rangle}\;,
\eer

\ber
&&\label{R4} (C^{1})_{a}:=\D^{b}\sigma_{ab}-\sfrac{2}{3}\tl\nb_{a}\Theta+\ep_{abc}\tl\nb^{b}\omega^{c}+q_{a}=0\;,\\
&&\label{R6} (C^{2})_{ a b}:=\ep_{cd(a}\tl\nb^{c}\sigma_{b)}{}^{d}+\tl\nb_{\langle a}\omega_{b \rangle}-H_{a b}=0\;,\\
&&\label{B6} (C^{3})_{a}:=\tl\nb^{b}H_{ab}+(\mu+p)\omega_{a}+\sfrac12\ep_{abc}\tl\nb^{b}q^{c}=0\;,\\
&&\label{B5} (C^{4})_{a}:=\tl\nb^{b}E_{ab}+\sfrac{1}{2}\tl\nb^{b}\pi_{ab}-\sfrac13\tl\nb_{a}\mu+
\sfrac13\Theta q_{a}=0\;,\\
&& \label{R5} (C^{5}):=\tl\nb^a\omega_a=0\;,\\
&&\label{B3} (C^{6})_{a}:=\tl\nb_{a}p_{m} +(\mu_{m}+p_{m}) A_{a}=0\;.
\eer
Eqns \eref{mue}-\eref{gmp} uniquely determine the covariant variables on some initial  (at $t=t_{0}$) hypersurface $S_{0}$ whereas Eqns \eref{R4}-\eref{B3} put restrictions on the initial data to be specified and must remain satisfied on any hypersurface $S_{t}$ for all comoving time $t\;.$ 

\subsection{The anti-Newtonian Limit}\label{antisec}

Anti-Newtonian universes are irrotational dust spacetimes \cite{maartens98} characterized by :
\ber
&&\label{C1an}p_{m}=0\;, ~~~~A_{a}=0\;,~~~q^{m}_{a}=0\;,~~~~\pi^{m}_{ab}=0\;,\nn
&&\label{Can}\om_{a}=0\;,~~~E_{ab}=0\;.
\eer
With these conditions,  the evolution equations \eref{mue}-\eref{gmp} in the anti-Newtonian regime can be rewritten as:
\ber
&&\label{mue2}\dot{\mu}_{m}=-\mu_{m}\Theta\;,\\
&&\label{murdot}\dot{\mu}_{R}=-(\mu_{R}+p_{R})\Theta+\frac{\mu_{m}f''}{f'^{2}}\dot{R}-\D^{a}q^{R}_{a}\;,\\
&&\label{ray2}\dot{\Theta}=-\sfrac13 \Theta^2-\sfrac12(\mu+3p)\;,\\
&&\label{qardot}\dot{q}^{R}_{a}=-\sfrac{4}{3}\Theta q^{R}_{a}+\frac{\mu_{m}f''}{f'^{2}}\D_{a}R-\D_{a}p_{R}-\D^{b}\pi^{R}_{ab}\;,\\
&&\dot{\sigma}_{ab}=-\sfrac{2}{3}\Theta\sigma_{ab}+\sfrac{1}{2}\pi^{R}_{ab}\;,\\
&&\label{pidotan}\dot{\pi}^{R}_{ab}=2\ep_{cd\langle a}\tl\nb^{c}H_{b\rangle }^{d}-\left(\mu+p_{R}\right)\sigma_{ab}-
\tl\nb_{\langle a}q^{R}_{b\rangle}-\sfrac{1}{3}\Theta\pi^{R}_{ab}\;,\\
\label{pidot}
&&\label{gmpan}\dot{H}_{ab} =-\Theta H_{ab}+
\sfrac{1}{2}\ep_{cd\langle a}\tl\nb^{c}\pi^{R~d}_{b\rangle }\;,
\eer

and are constrained by the following equations:

\ber
&& \label{c1an} (C^{1\ast })_{a}:=\D^{b}\sigma_{ab}-\sfrac{2}{3}\tl\nb_{a}\Theta+q^{R}_{a}=0\;,\\
&&\label{c2an} (C^{2\ast})_{a b}:=\ep_{cd(a}\tl\nb^{c}\sigma_{b)}{}^{d}-H_{a b}=0\;,\\
&&\label{c3an} (C^{3\ast})_{a}:=\tl\nb^{b}H_{a b}+\sfrac12\ep_{abc}\tl\nb^{b}q^{c}_{R}=0\;,\\
&&\label{c4an} (C^{4\ast})_{a}:=\tl\nb^{b}\pi^{R}_{ab}-\sfrac23\tl\nb_{a}\mu+
\sfrac23\Theta q^{R}_{a}=0\;.
\eer
An interesting aspect of the evolution equations \eref{mue2}-\eref{pidot} in the general relativistic treatment is that the propagation equations decouple from the gradient, divergence and curl terms (the spatial derivatives, basically), thus forming ordinary differential evolution equations. In other words, anti-Newtonian cosmologies in GR are {\it silent} models because  the flowlines emerging from the initial hypersurface $S_{0}$ evolve separately from each other \cite{elst97, maartens98, wyll06}. Looking at  Eqns \eref{murdot}, \eref{qardot}, \eref{pidotan} and \eref{gmpan},  the non-vanishing of the total anisotropic pressure and total heat flux (the presence of the components $\pi^{R}_{ab}$ and $q^{R}_{a}$) due to the addition of higher-order terms in $f(R)$ prompts one to  conclude that  \emph{anti-Newtonian models in $f(R)$ gravity are not silent}. However, using the linearized  identities \eref{C2}, \eref{C1}, Eqn\eref{c1an} and  Eqns \eref{qar}, \eref{pir} for  the definitions of $q^{R}_{a}$ and  $\pi^{R}_{ab}$, we can show that the curls vanish from Eqns \eref{gmpan} and \eref{c3an}:
\ber\label{curlq}
\ep^{acb}\tl\nb_c q^{R}_{b}&&=-\frac{1}{f'}\ep^{acb}\tl\nb_c\left[\left(f'''\dot{R}-\sfrac{1}{3}\Theta\right)\tilde{\nabla}_{b}R +f''\tilde{\nabla}_{b}\dot{R}\right]\nn
&&=-\frac{1}{f'}\left[\left(f'''\dot{R}-\sfrac{1}{3}\Theta\right)\ep^{acb}\tl\nb_c\D_{b}R+f''\ep^{acb}\tl\nb_c\D_{b}\dot{R} \right]\nn
&&=-\frac{1}{f'}\left[2\left(f'''\dot{R}-\sfrac{1}{3}\Theta\right)\dot{R}\om^{a}+2f''\ddot{R}\om^{a} \right]\nn
&&=-\frac{1}{f'}\om^{a}\left[2\left(f'''\dot{R}-\sfrac{1}{3}\Theta\right)\dot{R}+2f''\ddot{R} \right]=0\;,
\eer
\ber
\ep_{cd\langle a}\tl\nb^{c}\pi^{R~d}_{b\rangle }&&=\frac{f''}{f'}\ep^{acb}\tl\nb_c \tl\nb^d\left[\tl\nb_{\langle b}\tl\nb_{d \rangle}R-\dot{R}\sigma_{bd}\right]=\frac{f''}{f'}\ep^{acb}\tl\nb_c \left[\tl\nb^d\tl\nb_{\langle b}\tl\nb_{d \rangle}R-\dot{R}\tl\nb^d\sigma_{bd}\right]\nn
&&=\frac{f''}{f'}\ep^{acb}\left[\frac{2}{3}\D_{c}\D_{b}\D^{2}R+\sfrac{1}{3}\tl{R}\D_{c}\D_{b}R-\dot{R}\D_{c}\left(\sfrac{2}{3}\D_{b}\Theta-q^{R}_{b}\right)\right]\nn
&&=\frac{f''}{f'}\ep^{acb}\left[\sfrac{2}{3}\D_{c}\D_{b}\D^{2}R+\sfrac{1}{3}\tl{R}\D_{c}\D_{b}R-\frac{2}{3}\dot{R}\D_{c}\D_{b}\Theta\right]\nn
&&=\frac{f''}{f'}\left[\sfrac{2}{3}\tl{R}\dot{R}\om^{a}-\sfrac{4}{3}\dot{R}\dot{\Theta}\om^{a}\right]=\frac{f''}{f'}\om^{a}\left[\sfrac{2}{3}\tl{R}\dot{R}-\sfrac{4}{3}\dot{R}\dot{\Theta}\right]=0\label{curlpi}
\eer
since an irrotational dust fluid is assumed by construction.

Thus we have  simplified evolution and constraint equations for the Weyl tensor
\ber
&&\label{gmpan2}\dot{H}_{ab} =-\Theta H_{ab}\;,\\
&&(C^{3\ast\ast})_{a}:=\tl\nb^{b}H_{a b}=0\;.
\label{c3an2}
\eer

 Also important to note is that there are no new constraint  equations arising from the  GE-free assumption, contrary to the GR result where \eref{pidotan} is the constraint equation (with $\pi^{R}_{ab}=0=\dot{\pi}^{R}_{ab}$) from which the integrability conditions arise.
\section{Integrability Conditions}\label{intsec}
Eqn \eref{c4an}  is a modified constraint due to the vanishing of  $E_{ab}$  from Eqn \eref{B5}. Since the consistency of the ``new'' set of constraint equations \eref{c1an}, \eref{c2an}, \eref{c4an} and  \eref{c3an2} is not {\it a priori} guaranteed under this GE-free restriction, we need to show that the initial conditions on $S_{0}$ are consistent, and that these constraint equations are preserved under evolution on $S_{t}$.

\subsection{Spatial consistency}

To check the spatial consistency of the 
 constraint in Eqn \eref{c4an} on any $S_{0}$, let us take the curl of this equation:
 \ber
&&0=\frac{f''}{f'}\ep^{acb}\tl\nb_c \tl\nb^d\pi^{R}_{bd}-\sfrac{2}{3}\ep^{acb}\tl\nb_c\tl\nb_b\mu+\sfrac{2}{3}\Theta \ep^{acb}\tl\nb_c q^{R}_{b}\nn
&&=\om^{a}\bigg\{\frac{f''}{f'}\left[\sfrac{2}{3}\tl{R}\dot{R}-\sfrac{4}{3}\dot{R}\dot{\Theta}\right]-\sfrac{4}{3}\dot{\mu}-\left(\frac{\dot{R}f''}{f'^{2}}+\frac{2\Theta}{3f'}\right)\left[2\left(f'''\dot{R}-\sfrac{1}{3}\Theta\right)\dot{R}+2f''\ddot{R} \right]\bigg\}\nn
&&=0\;,
\label{spacon}
\eer
where the results of Eqns \eref{curlq}-\eref{curlpi} have been used.
 Thus \emph{the constraints are consistent with each other.}

\subsection{ Temporal consistency}

Let us now check if the constraint equations are preserved under evolution. Evolving \eref{c4an} gives
\ber
0&&=\left(\tl\nb^{b}\pi^{R}_{ab}\right)^{.}-\sfrac23\left(\tl\nb_{a}\mu\right)^{.}+\sfrac23\dot{\Theta} q^{R}_{a}+\sfrac23\Theta \dot{q}^{R}_{a}\nn
&&=\tl\nb^{b}\dot{\pi}^{R}_{ab}-\sfrac{1}{3}\Theta\D^{b}\pi^{R}_{ab}-\sfrac23\left(\tl\nb_{a}\dot{\mu}-\sfrac{1}{3}\Theta\D_{a}\mu\right)+\sfrac23\dot{\Theta} q^{R}_{a}+\sfrac23\Theta \dot{q}^{R}_{a}\;,\eer
where in the second step we have used the commutation relations \eref{a15} and \eref{a14}. We can expand the  RHS of the last equation using Eqn \eref{pidot}:
\ber
0&&=\tl\nb^{b}\left[2\ep_{cd\langle a}\tl\nb^{c}H_{b\rangle }^{d}-\left(\mu+p_{R}\right)\sigma_{ab}-
\tl\nb_{\langle a}q_{b\rangle}{}^{R}-\sfrac{1}{3}\Theta\pi^{R}_{ab}\right]\nn
&&~~~-\sfrac{1}{3}\Theta\D^{b}\pi^{R}_{ab}-\sfrac23\left(\tl\nb_{a}\dot{\mu}-\sfrac{1}{3}\Theta\D_{a}\mu\right)+\sfrac23\dot{\Theta} q^{R}_{a}+\sfrac23\Theta \dot{q}^{R}_{a}\nn
&&=\ep_{abc}\tl\nb^{b}\tl\nb_{d}H^{cd}-\left(\mu+p_{R}\right)\tl\nb^{b}\sigma_{ab}-\tl\nb^{b}
\tl\nb_{\langle a}q_{b\rangle}{}^{R}\nn
&&~~~-\sfrac{2}{3}\Theta\D^{b}\pi^{R}_{ab}-\sfrac23\tl\nb_{a}\dot{\mu}+\sfrac{2}{9}\Theta\D_{a}\mu+\sfrac23\dot{\Theta} q^{R}_{a}+\sfrac23\Theta \dot{q}^{R}_{a}\;.\label{tempcon}
\eer
Using Eqns \eref{c3an2}, \eref{c1an} and \eref{c4an} for the divergences of $H_{cd}$, $\sigma_{ab}$ and $\pi^{R}_{ab}$ and making use of the linearized  vectorial identity \cite{carloni08} 
\be
\tl\nb^{b}\tl\nb_{\langle a}V_{b\rangle}=\sfrac{1}{2}\D^{2}V_{a}+\sfrac{1}{6}\D_{a}(\D^{b}V_{b})+\sfrac{1}{6}\tl{R}V_{a}\;,
\ee 
we can rewrite the RHS of Eqn \eref{tempcon} as
\ber
&&-\left(\mu+p_{R}\right)\left(\frac{2}{3}\D_{a}\Theta-q^{R}_{a}\right)-\sfrac{1}{2}\D^{2}q^{R}_{a}-\sfrac{1}{6}\D_{a}(\D^{b}q^{R}_{b})-\sfrac{1}{6}\tl{R}q^{R}_{a}\nn
&&-\sfrac{4}{9}\Theta\D_{a}\mu+\sfrac{4}{9}\Theta^{2}q^{R}_{a}+\sfrac23\nb_{a}\left[\left(\mu+p_{R}\right)\Theta+\D^{a}q^{R}_{a}\right]+\sfrac{2}{9}\Theta\D_{a}\mu+\sfrac23\dot{\Theta} q^{R}_{a}+\sfrac23\Theta \dot{q}^{R}_{a}\nn
&&=-\left(\mu+p_{R}\right)\left(\frac{2}{3}\D_{a}\Theta-q^{R}_{a}\right)-\sfrac{1}{2}\D^{2}q^{R}_{a}-\sfrac{1}{6}\D_{a}(\D^{b}q^{R}_{b})-\sfrac{1}{6}\tl{R}q^{R}_{a}-\sfrac{4}{9}\Theta\D_{a}\mu+\sfrac{4}{9}\Theta^{2}q^{R}_{a}\nn
&&+\sfrac{2}{3}\D^{2}q^{R}_{a}+\sfrac23\Theta\tl\nb_{a}\mu+\sfrac23\Theta\tl\nb_{a}p_{R}+\sfrac23(\mu+p_{R})\D_{a}\Theta+\sfrac{2}{9}\Theta\D_{a}\mu+\sfrac23\dot{\Theta} q^{R}_{a}+\sfrac23\Theta \dot{q}^{R}_{a}\nn
&&=\sfrac23\Theta \dot{q}^{R}_{a}-\sfrac{1}{6}\D_{a}(\D^{b}q^{R}_{b})+\sfrac{1}{6}\D^{2}q^{R}_{a}+\sfrac{1}{3}\left(\mu+\Theta^{2}\right)q^{R}_{a}+\sfrac23\Theta\tl\nb_{a}p_{R}+\sfrac{4}{9}\Theta\D_{a}\mu\;.
\eer
Thus a consistent evolution of the constraints requires that
\be\label{concons0}
\Theta \dot{q}^{R}_{a}-\sfrac{1}{4}\D_{a}(\D^{b}q^{R}_{b})+\sfrac{1}{4}\D^{2}q^{R}_{a}+\sfrac{1}{2}\left(\mu+\Theta^{2}\right)q^{R}_{a}+\Theta\tl\nb_{a}p_{R}+\sfrac{2}{3}\Theta\D_{a}\mu=0\;.
\ee
Since one can use  Eqn \eref{qardot} to substitute for $\dot{q}^{R}_{a}\;,$ Eqn \eref{concons}  can be further simplified and rewritten as
\be\label{concons}
\D^{2}q^{R}_{a}-\D_{a}(\D^{b}q^{R}_{b})+\tl{R}q^{R}_{a}+\frac{4f''}{f'^{2}}\mu_{m}\Theta\D_{a}R=0\;.
\ee
Let us take a look at the implications of this equation in two cases.
\begin{enumerate}
\item \underline{Flat universes ($K=0=\tl{R}$):}
In this case, the condition \eref{concons} together with \eref{a19} holds only if 
\be
\frac{f''}{f'^{2}}\mu_{m}\Theta\D_{a}R=0\;,
\ee
but since we are dealing with a dust universe in fourth-order gravity, we impose $\mu_{m}\neq 0$ and $f''\neq 0$.  Thus \emph {for a consistently evolving set of constraints in the flat, anti-Newtonian spacetimes, either $\Theta=0$ (``static'') or $\D_{a}R=0\;.$}  
\item \underline{Closed \& Open universes ($K=\pm1$):}
\emph{Any dust solution of 
\be
\frac{1}{f'}\bigg\{\left[\frac{f''\mu_{m}\Theta}{f'}\mp\frac{2}{a^{2}}\left(\dot{R}f'''-\sfrac{1}{3}\Theta f''\right)\right]\D_{a}R\mp\frac{2f''}{a^{2}}\D_{a}\dot{R}\bigg\}=0\;
\ee 
with $ f''\neq 0$ is an anti-Newtonian solution.}
\end{enumerate}
\section{Perturbation equations}\label{pertsec}

The usual covariant and gauge-invariant inhomogeneity variables of standard matter and expansion are given by \cite{DBE92, EB89}

\ber
&&\md^{m}_{a}=\frac{a\D_{a}\mu_{m}}{\mu_{m}}\;,~~~~~~~\mz_{a}=a\D_{a}\Theta\;,\end{eqnarray}
whereas the information about our deviation from standard GR is carried by the following  dimensionless gradient quantities  \cite{carloni08, carlonireview}:
\be
{\car}_{a}=a\D_{a}R\;,~~~~~~ \Re_{a}=a\D_{a}\dot{R}\;.
\ee
Using these perturbation variables, one can rewrite Eqn \eref{concons} as
\be\label{pertcons}
\frac{1}{f'}\bigg\{\left[\frac{f''\mu_{m}\Theta}{f'}-\frac{2K}{a^{2}}\left(\dot{R}f'''-\sfrac{1}{3}\Theta f''\right)\right]{\car}_{a}-\frac{2Kf''}{a^{2}}\Re_{a}\bigg\}=0\;.
\ee 
For a general $f(R)$ gravity, the complete set of linearized evolution equations of cosmological perturbations in a pure dust cosmic medium  is given by the 4 first-order coupled system of equations \cite{abebe12, carloni08}:
\ber\label{deldota}
&&\dot{\md}^{m}_{a}+\mz_{a}=0\;,\\
&&\label{zdota} \dot{\mz}_{a}-\left(\dot{R}\frac{f''}{f'}-\frac{2}{3}\Theta\right)\mz_{a}+\frac{\mu_{m}}{f'}\md^{m}_{a}-\Theta\frac{f''}{f'}\Re_{a}+\frac{f''}{f'}\tl \nb^{2}{\car}_{a}\nn
&&-\left( \frac{1}{2}-\frac{1}{2}\frac{ff''}{f'^{2}}+\frac{f''\mu_{m}}{f'^{2}}-\dot{R}\Theta(\frac{f''}{f'})^{2}+\dot{R}\Theta \frac{f'''}{f'}+\frac{2K}{a^{2}}\frac{f''}{f'}\right) {\car}_{a}=0\;,\\
&&\label{cardota}\dot{\car}_{a}-\Re_{a}=0\;,\\
&&\label{redota}\dot{\Re}_{a}+\left(\Theta +2\dot{R}\frac{f'''}{f''}\right)\Re_{a}+\dot{R}\mz_{a}-\frac{\mu_{m}}{3f''}\md^{m}_{a}-\tl\nb^{2}\car_{a}\nn
&&+\left(\ddot{R}\frac{f'''}{f''}+\dot{R}^{2}\frac{f^{(4)}}{f''}+\Theta\dot{R}\frac{f'''}{f''}+\frac{f'}{3f''}-\frac{R}{3}+\frac{2K}{a^{2}}\right)\car_{a}=0\;.
\eer

From the continuity equation \eref{mue2}, we see that a  static dust spacetime implies $\mu_{m}\neq \mu_{m}(t)\;.$ Using  Eqn \eref{ray2}  together with the trace equation 
\be\label{trace}
3f''\ddot{R}+3\dot{R}^2f'''+3\Theta\dot{R}f''-3f''\D^2R-\mu_m-Rf'+2f=0\;,
\ee
one can then show that the density and curvature perturbations are related by
\be
\mu_{m}\md^{m}_{a}-\sfrac{1}{2}f'{\car}_{a}+f''\D^{2}{\car}_{a}=0\;,
\ee
whereas the evolution equations in flat spacetime would reduce to the system
\ber\label{deldotas}
&&\dot{\md}^{m}_{a}=0\;,\\
&&\label{cardotas}\dot{\car}_{a}-\Re_{a}=0\;,\\
&&\label{redotas}\dot{\Re}_{a}+2\dot{R}\frac{f'''}{f''}\Re_{a}+\left(\ddot{R}\frac{f'''}{f''}+\dot{R}^{2}\frac{f^{(4)}}{f''}+\frac{f'}{6f''}-\frac{R}{3}\right)\car_{a}-\sfrac{2}{3}\tl\nb^{2}\car_{a}=0\;.
\eer
Thus $\md^{m}_{a}$ is constant in time and the last two equations reveal that the evolution of curvature inhomogeneities is unaffected by the matter perturbations.

The trace equation ensures that
 the condition $\D_{a}R=0 ~(\implies R=R(t))$ does not necessarily lead to $(\mu_{m}=\mu_{m}(t)\implies) \D_{a}\mu_{m}=0\;.$
Hence for the flat non-static  anti-Newtonian solution, it can be shown that the above set of perturbation equations \eref{deldota}-\eref{redota} can be reduced to two decoupled first-order evolution equations:
\ber\label{dotdan}
&&\dot{\md}^{m}_{a}+\frac{\mu_{m}}{3\dot{R}f''}\md_{a}=0\;,\\
&& \dot{\mz}_{a}+2\left(\dot{R}\frac{f''}{f'}+\frac{1}{3}\Theta\right)\mz_{a}=0\;.\label{dotzan}
\eer
This is an interesting, and perhaps a surprising, result showing the growth of structure unaffected by the expansion, and vice versa.

In the general (non-flat) case, using \eref{cardota} to substitute for $\Re_{a}$ in \eref{pertcons} results in an evolution equation for ${\car}_{a}$ decoupled from matter and expansion perturbations at first order: 
\be\label{pertcons2}
\dot{\car}_{a}+\left(\frac{\dot{R}f'''}{f''}-\sfrac{1}{3}\Theta -\frac{a^{2}\mu_{m}\Theta}{2Kf'}\right){\car}_{a}=0\;.
\ee 
We note that,  given a class of $f(R)$ gravity with a known background expansion, the perturbation equations obtained in the limiting anti-Newtonian cases can be solved with much more ease than in the general case. It will be interesting to look for cosmologically viable $f(R)$ models  (models that mimic early universe inflation, transient radiation- and matter-dominated epochs and late-time accelerated expansion) that also satisfy the anti-Newtonian solutions above. A full investigation of such models in the context of large-scale power-spectrum and gravitational collapse will be addressed in a future work.

\section{Results and Discussion}\label{concsec}

In this paper,  a linearized covariant consistency analysis of dust universes with vanishing gravito-electric part of the Weyl tensor has been presented for $f(R)$ gravity. We have pointed out that, unlike their  GR counterparts, these  anti-Newtonian models are generally not silent. We have also shown the existence of an integrability condition for generic $f(R)$ gravity models. 

The perturbation equations resulting from the integrability condition imposed have been presented. For flat static models, we have shown that the dust perturbations are constant, and the ``curvature perturbations'' ({\it i.e.} ${\car}_{a}$ and $\Re_{a}$) are decoupled from the matter fluctuations at first order. For flat non-static models, it has been shown that the density contrast and the expansion evolve unhindered by each other.

Likewise, we have shown that the fluctuations in the Ricci curvature scalar evolve independently of the matter and expansion gradients in the open and closed class of anti-Newtonian models.

 A  power-spectrum analysis of these anti-Newtonian solutions for some cosmologically viable classes of $f(R)$ models as well as a covariant analysis  for general nonlinear models and their integrability conditions is left for a subsequent work.
\ack The author thanks Rodney Medupe and Rituparno Goswami for comments and suggestions and acknowledges the hospitality of the Astrophysics Group (Department of Physics, NWU-Mafikeng) during the preparation of this work. 
\appendix
\section{Useful linearized differential identities \cite{maartens98, maartens97, carloni08}}
The following linearized identities hold for all vectors and tensors that vanish in the background,
$S_{ab}=S_{\la ab\ra}$. (Nonlinear identities are given
in \cite{maart97,hve96,maar97}.) 
\begin{eqnarray}
\ep^{abc}\D_b \D_cf &=&-2\dot{f}\omega_a \,,
\label{a13} \\
\D^2\left(\D_af\right) &=&\D_a\left(\D^2f\right) 
+\sfrac{1}{3}\tl{R}\D_a f+2\dot{f}\ep_{abc}\D^b\omega^c
\,, \label{a19}\\
\left(\D_af\right)^{\rd} &=& \D_a\dot{f}-\sfrac{1}{3}\Theta\D_af+\dot{f}A_a \,,
\label{a14}\\
\left(\D_aS_{b\cdots}\right)^{\rd} &=& \D_a\dot{S}_{b\cdots}
-\sfrac{1}{3}\Theta\D_aS_{b\cdots}\,,
\label{a15}\\
\left(\D^2 f\right)^{\rd} &=& \D^2\dot{f}-\sfrac{2}{3}\Theta\D^2 f 
+\dot{f}\D^a A_a \,,\label{a21}\\
\D_{[a}\D_{b]}V_c &=& 
-\sfrac{1}{6}\tl{R}V_{[a}h_{b]c}
\,, \label{a16}\\
\D_{[a}\D_{b]}S^{cd} &=& -\sfrac{1}{3}\tl{R}S_{[a}{}^{(c}h_{b]}{}^{d)} \,, \label{a17}\\
\D^a\left(\ep_{abc}\D^bV^c\right) &=& 0\;, \label{a20}\\
\label{divcurl}\D_b\left(\ep^{cd\la a}\D_c S^{b\ra}_d\right) &=& {\ts{1\over2}}\ep^{abc}\D_b \left(\D_d S^d_c\right)\,.
\label{a18}
\end{eqnarray}
%
\section*{References}

\bibliography{@bibliography}
\bibliographystyle{iopart-num}

\end{document}